# Machine-Learning-Assisted Metasurface Design for High-Efficiency Thermal Emitter Optimization

Zhaxylyk A. Kudyshev[1,2], Alexander V. Kildishev[1],
Vladimir M. Shalaev[1], and Alexandra Boltasseva[1]

[1] School of Electrical and Computer Engineering and Birck Nanotechnology Center, Purdue University, West Lafayette, IN, 47906, USA

[2] Center for Science of Information, Purdue University, West Lafayette, IN, 47906, USA

**Abstract:** With the emergence of new photonic and plasmonic materials with optimized properties as well as advanced nanofabrication techniques, nanophotonic devices are now capable of providing solutions to global challenges in energy conversion, information technologies, chemical/biological sensing, space exploration, quantum computing, and secure communication. Addressing grand challenges poses inherently complex, multi-disciplinary problems with a manifold of stringent constraints in conjunction with the required system's performance. Conventional optimization techniques have long been utilized as powerful tools to address multi-constrained design tasks. One example is so-called topology optimization that has emerged as a highly successful architect for the advanced design of non-intuitive photonic structures. Despite many advantages, this technique requires substantial computational resources and thus has very limited applicability to highly constrained optimization problems within high-dimensions parametric space. In our approach, we merge the topology optimization method with deep learning algorithms such as adversarial autoencoders and show substantial improvement of the optimization process in terms of computational time (4900 times faster) and final devices efficiencies (~98%), by providing unparalleled control of the compact design space representations. By enabling efficient, global optimization searches within complex landscapes, the proposed compact hyperparametric representations could become crucial for multi-constrained problems. The proposed approach could enable a much broader scope of the optimal designs and data-driven materials synthesis that goes beyond photonic and optoelectronic applications.

# I. INTRODUCTION

Realization of practical optical structures and devices is an inherently complex problem due to multi-faceted requirements with the manifold of stringent constraints on the optical performance, materials, scalability, and experimental tolerances. These multiple requirements inevitably open up an enormously large optimization space. Despite the complexity of the available parametric space, almost all nanophotonic structures up to date are designed either intuitively or based on a priori selected topologies, and by adjusting a very limited number of parameters (e.g., the periodicity, the trivial geometrical shapes, and dimensions of the resonant elements). Such intuition-based models are only useful for *ad hoc* needs and have limited applicability and predictive power. The exhaustive parameter sweeps are often done "by hand." Since the comprehensive search in hyper-dimensional design space is highly resource-heavy, multi-objective optimization has so far been almost impossible. Moreover, human's restrained capacity to think hyper-dimensionally limits our perception of multivariate optimization models. Thus, advanced machinery is needed to manage the multi-domain, hyper-dimensional design parameter space.

The innovatory field of the inverse design has recently been transforming conventional nanophotonics and allowing for the discovery of unorthodox optical structures via computer algorithms rather than engineered "by hand"[1]. The realization of 'non-intuitive' designs requires truly new approaches combined with already established diverse optimization and sensitivity methods such as genetic algorithms[2–4] and different variations of the adjoint method[5–12]. Particularly, topology optimization (TO) that previously revolutionized mechanical and aerospace engineering[13–15] by providing unexpected solutions to constrained material distribution problems, has recently emerged as a powerful architect for photonic design[9,16–20] that offers broader parameter space and flexible incorporation with different computational methods. However, the gradient descent nature of the TO method, in which an initial device design is locally optimized within the parametric space, significantly relies on the initial guess of the material distribution inside the optimization domain. Lack of intuition in choosing the right initial geometries leads to multiple TO runs in order to select the best-performing solution. Since TO is very computationally expensive, this substantially limits its applicability to multi-constrained problems that require a significant expansion of the optimization space to larger parametric domains that include mechanical, chemical, and optical properties.

Recently, different aspects of machine learning (ML) have attracted major interest in the field of nanophotonics[21–25]. Various discriminative deep learning models have been adapted to find the solution to direct and inverse electrodynamics problems[26–29]. Unlike conventional electrodynamic simulation methods, which require intensive, time-consuming computations, ML algorithms enable almost instantaneous solution searches due to the learning process performed during the training phase. Along with the pure discriminative model, various generative networks, such as generative adversarial networks (GANs)[30] and variation autoencoders (VAEs)[31], have been used for nanophotonic design optimization. Recently, GANs have been coupled with the TO method for optimizing diffractive dielectric gratings[32,33]. It has been shown that adversarial networks could be efficiently trained on topology optimized designs for the rapid generation of large families of highly efficient grating designs in the significantly smaller timescale.

Within this work, we demonstrate that adversarial autoencoders (AAE) can be efficiently adapted for rapid nanophotonic design optimization. Mainly, we show that AAE coupled with a TO engine (i) *enables >4900 times faster optimization search within the compressed design space* (latent space), and moreover, (ii) *ensures unparalleled control over the latent space distribution*. The latter is essential for multi-constrained optimization problems, where a compact hyper-parametric representation becomes critical for efficient optimization searches within a complex landscape. To showcase our AAE assisted method, we optimize a metasurface thermal emitter for thermophotovoltaic (TPV) applications. Compared to an adjoint-based TO design with 92% efficiency of the thermal emission reshaping, the proposed method provides three times speedup and gives 98% efficiency. The proposed approach can be adapted to a broader scope of the problems in optics, chemistry, and mechanics.

## II. ADVERSARIAL AUTOENCODERS FOR DESIGN OPTIMIZATION

Generative models raised significant interest in the machine learning community due to a fundamentally new approach to data interpretation. Generative networks aim to learn dataset distribution of the training set and generate new data with some additional variations, unlike discriminative models, which learn hard/soft boundaries between classes of the data.

Within this work, we coupled AAE generative network with conventional TO method for highly efficient topology optimization of nanophotonic devices. The chosen AAE consists of three coupler neural networks: encoder, decoder/generator, and discriminator (Fig.1), following an initial AAE concept [34]. Similar to variation autoencoders (VAEs)[35], the main goal of the encoder in AAE is to compress a given input pattern into a compact, continuous design space, so-called latent space. Then, the decoder learns to reconstruct the real space patterns based on a given latent space coordinate – latent vector. In contrast to VAEs, where the latent space distribution is assumed to be a standard normal distribution, AAE performs adversarial learning (like in GANs[36]) by applying the discriminator to force latent space to pre-defined model distribution. AAE can be considered as a combination of VAE and GAN networks.

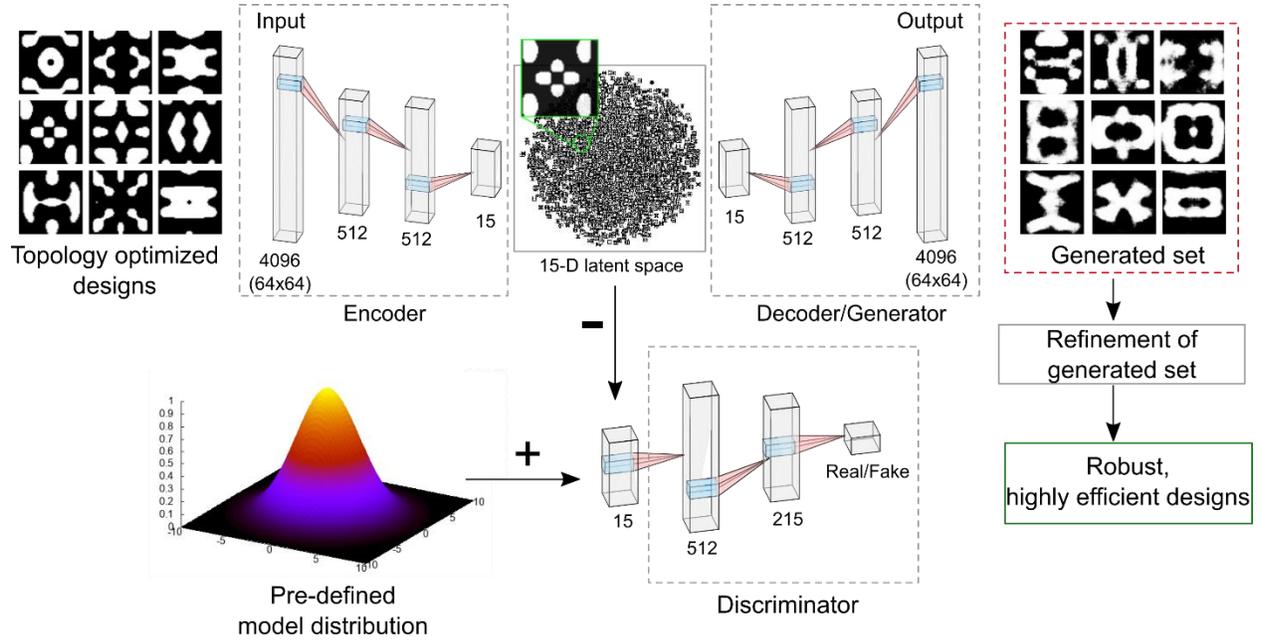

**Figure 1. AAE-assisted topology optimization.** Starting from a discrete set of the topology optimized designs of nanoantenna that serves as the metasurface building block, the encoder compresses each nanoantenna design into a point in the latent space – compressed, continuous design space. The decoder reconstructs the design based on the input coordinate in the latent space. The discriminator forces the encoder to construct the latent space with a pre-defined distribution. The trained decoder is then used as a generator, which takes the latent space coordinate as an input and generates a large set of designs. The structure refinement procedure is applied to the generated set to eliminate unstable, low-efficient designs.

Figure 1 shows the proposed flow of the optimization process. AAE assisted optimization consists of the three main steps: (i) data generation using adjoint TO method; (ii) training AAE network on the generated dataset; (iii) structure refinement.

The first step aims at generating efficient designs using the TO and construction of an appropriate training dataset. In the second step, the AAE network is trained on the constructed dataset of TO designs. During the training process, the encoder is forced to produce (i) a latent space that the decoder can use for reconstruction, and (ii) *a latent sampling that could pass the discriminator as a sample from the pre-defined model distribution*. Once the AAE network is trained, then the decoder can be used as a generator that takes the latent vector as an input and generates a design pattern in the real space. In the third step, a design refinement procedure filters out unstable, low-efficient designs.

The choice of the AAE network over other deep generative network architectures, such as VAE and GANs, is motivated by its important advantages. First, the AAE method provides the neural network with *the dense (continuous) latent space representation*[34] that becomes critical for interpolating the hyper-dimensional parametric space. Such a continuous representation enables a much broader variety of the generated designs. Secondly, the AAE approach sets no specific limits on the pre-defined model distributions, hence enabling extremely flexible control over the latent space configuration[34]. Finally, the AAE networks have better trainability in comparison with GANs[37] because the discriminator in the AAE networks is applied to a compressed continuous latent representation in comparison with GANs, where adversarial learning is applied directly to the patterns/images. Details on the quantitative comparison between GAN and AAE networks can be found in Supplementary Materials. Within this work, we showcase the proposed approach by optimizing a metasurface thermal emitter design for TPV applications. The next section highlights the main constraints of the problem.

## III. THERMAL EMITTER DESIGN CONSTRAINTS

The conventional TPV engine aims at generating electrical power by radiative heat transfer and usually consists of a heater and a photovoltaic (PV) cell array (Fig. 2a). Without losing generality, we consider TPV systems utilizing GaSb PV cells with a working band ranging from $\lambda_{min} = 0.5 \mu m$ to $\lambda_{max} = 1.7 \mu m$. To ensure efficient electrical power generation, thermal emission of the heater should significantly overlap with the working band of the PV cell. Hence, the temperature of the heater should exceed 1000 ˚C. Figure 1b shows the emission spectrum of the blackbody at 1800 ˚C. However, even for the appropriately high temperature, only a small portion of the emission overlaps with the PV cell working band (Fig. 2b, green area), while most of the emission energy remains outside the band (Fig. 2b, red area). While in-band radiation produces electron-hole pairs, out-of-band radiation causes undesirable heating of the PV cell, significantly reducing its quantum efficiency and device lifetime. By patterning the surface of the heater with a properly designed thermal emitting metasurface it is possible to spectrally reshape the emissivity $\varepsilon(\lambda)$ of the heater to maximize the in-band and minimize the out-of-band radiation. The ideal thermal emitter has a step function type profile of the emissivity with $\varepsilon(\lambda_{min} \leq \lambda \leq \lambda_{max}) = 1$ and zero elsewhere (dashed blue contour, Fig.2b).

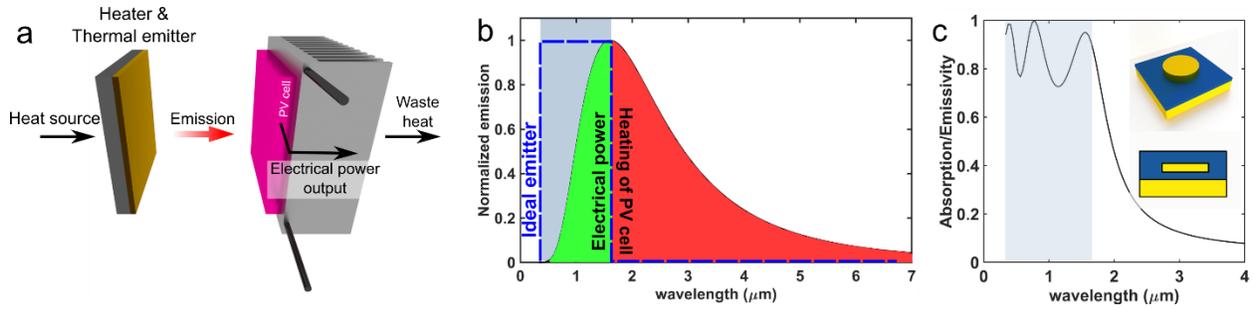

**Figure 2. Thermal emitter for TPV applications.** (a) Schematic of a TPV engine: a heater patterned with a thermal emitter array and a PV cell. (b) Blackbody radiation of the bare heater (solid black curve) corresponding to emission of blackbody at 1500 ˚C. The grey rectangular region highlights the GaSb PV cell working band. Only in-band radiation is converted into electrical power (green area), while out-of-band radiation cases heating of the PV cell (red area). Blue dashed contour corresponds to an ideal thermal emitter's emissivity/absorption spectrum. (c) Absorption/emissivity spectrum of the optimized cylindrical gap plasmon thermal emitter (shaded region corresponds to the GaSb PV cell's working band). The cylindrical emitter's unit cell size is 145 nm, radius and height of the cylinder are 50 nm and 30 nm respectively, the $Si_3N_4$ spacer is 40 nm thick, top $Si_3N_4$ cover is 90 nm thick. Inset shows the 3D and a side views of the structure

Due to high-temperature operation, thermally emitting metasurfaces should be designed utilizing high-temperature stable material platforms. Recently, it has been demonstrated that transition metal nitrides (TiN, ZrN) exhibit metal-like optical properties and plasmonic attributes on par with noble metals in the visible and near-infrared spectral regions[38–42]. In contrast to noble metals conventionally used in plasmonics, transition metal nitrides are stable at very high temperatures[43]. In this work, we focus on the TiN/$Si_3N_4$ material combination for metasurface thermal emitter designs [44]. More details on dielectric permittivity functions of TiN and $Si_3N_4$ in Supplementary Materials (Section S1).

Recently, various selective emitter designs have been investigated including rare-earth oxides[45], photonic crystals[46–48] and metamaterial/metasurface-based emitters[49,50]. One of the most commonly used designs of the thermal emitter is a gap plasmon structure[51–54], which consists of a back-reflector, dielectric spacing material, and a top array with plasmonic resonators of simple geometrical shapes[55]. This design offers simple fabrication, as well as intuitive design. However, the intuitive simple shapes substantially reduce the degrees of freedom for optimization; and as a result, significantly limit achievable efficiencies. As a reference, we use the parametric optimization of a gap plasmon[56] structure that comprises an array of TiN cylindrical resonators. TiN cylinders are deposited on top of a $Si_3N_4$ spacer layer that covers an optically thick TiN back reflector (see the inset in Fig. 2c). Optimization of the array period, dielectric spacer thickness as well as the dimensions of the TiN cylinder (radius and height) is performed with particle swarm optimization method[57], minimizing the norm difference between emissivity/absorption spectrum of the structure with the ideal emitter emissivity. Figure 2c shows the obtained absorption spectrum. While the out-of-band emissivity is substantially suppressed, the mean in-band emissivity/absorption reaches only 84% due to a limited number of resonant in-band modes. This reference case demonstrates that even though the parametric design space is large enough, the trivial initial shape of the resonator fundamentally limits the ultimate achievable efficiency.

As the next step, a material distribution within the simulation domain can be used as a sub-set of the design parameter space. Hence, by applying a gradient-based TO technique to such optimization domain it can converge to an optimal, non-intuitive binary material distribution that enables highly efficient device performance. In the next section, we adapt such a density-based TO technique to construct a training dataset for an AAE network for optimizing the metasurface thermal emitter design.

**IV. TO-GENERATED TRAINING SET AND LATENT SPACE ENGINEERING**

To generate a training set with topology optimization, we consider a gap plasmon metasurface configuration. The three-layer structure comprises an optically thick back TiN reflector, a 40-nm-thick $Si_3N_4$ spacer, and a 120-nm-thick top layer, with a fixed 280-nm-period of the unit cell in both lateral directions (Fig. 3a). The comparison of the TO of the thermal emitter with different unit cell sizes can be found in the Supplementary Materials. The top layer is defined as the optimization region and discretized with 60 × 60 optimization elements. Initially, the material distribution in the optimization region is set to be a random, smooth dielectric function within one quadrant of the unit cell, and is translated using mirror symmetry to the whole unit cell. Here we have set-up pre-defined symmetry properties along x and y directions, however, in more general cases setting up a random initial guess into the whole unit cell will lead to topology optimized structures with arbitrary symmetries. Interpolation of the material distribution is done using a non-linear interpolation scheme proposed in [58]. TO is realized using an adjoint optimization scheme, which requires two full-wave simulations per iteration (forward and adjoint) and employs a direct commercial full-wave solver built on finite-difference time-domain (FDTD) approximation of the Maxwell equations (Lumerical FDTD) controlled with a Matlab host script. The spatial distribution of the forward and adjoint fields determines the "heat map" of the dielectric function perturbation inside the optimization domain, which maximizes the figure of merit gradient at a given iteration. After multiple iterations, the material distribution inside the optimization region converges to a binary structure (air/TiN). One of the critical constraints of the optimization procedure is compatibility with available fabrication techniques, i.e., the stability of the final design to fabrication imperfections and achievable fabrication tolerances[6,59]. This constraint requires incorporating a two-step robustness algorithm into the optimization procedure: i) elimination of the sub-precision features by averaging the permittivity of the design over neighboring regions and pushing the design to binary structures with the next optimization evolution; ii) averaging the figure of merit over the perturbed geometries of the device, hence, reducing the impact of geometric variability on the device efficiency. Thus, we apply spatial filtering, which eliminates the sub-30 nm features from the designs in each 10$^{th}$ iteration of the optimization algorithm, while the total number of TO iterations is set to 50. More details on the implementation of the direct TO can be found in Supplementary Materials (Section S2).

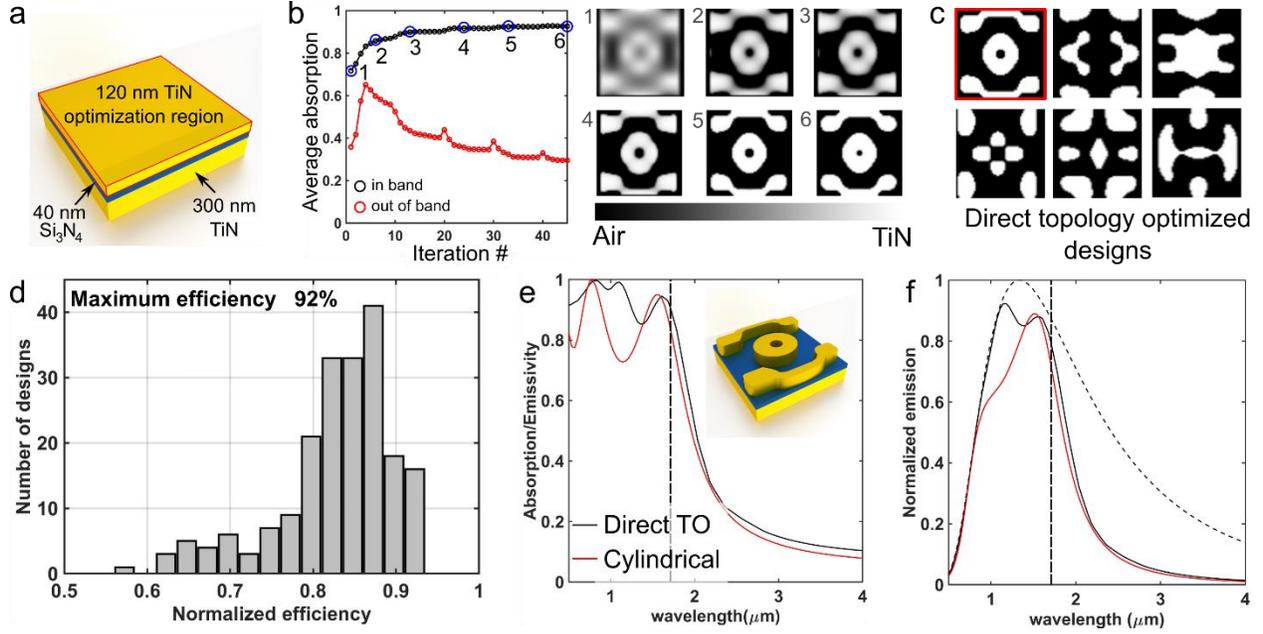

**Figure 3. Direct Topology Optimization for training dataset generation.** (a) The base structure under consideration consists of a 300-nm-thick TiN back reflector, 40-nm-thick $Si_3N_4$ dielectric spacer, and 120-nm-thick top TiN patterned layer in 280×280 $nm^2$ unit cell. The top TiN layer is set to be the optimization region (outlined by red box). (b) Convergence plot of the topology optimization with corresponding evolution of the material distribution. Black and red curves correspond to the in-band and out-of-band average absorption respectively. (c) Designs obtained by the direct topology optimization, including the design with the highest efficiency (92%) framed in the red box (black color corresponds to air, white to TiN). (d) Statistics of efficiency distribution of 200 topology optimized designs. Normalized efficiency is defined as a ratio of in-band emissions of the TO thermal emitter and of an ideal emitter at 1800 °C. (e) The absorption/emissivity spectrum of the best TO (black) and cylindrical (red) thermal emitters; (f) Corresponding emissivity spectra at 1800 °C: the best TO emitter (black), cylindrical emitter (red), and blackbody emission (dashed black curve).

For the TO process, the figure of merit (FOM) is defined as the spectrally weighted average of the in-band absorption and the out-of-band reflectivity. The weighting of the FOM spectrum is done with respect to the absorption and reflection spectrum of the ideal emitter (Supplementary Materials, section S2). The FOM corresponds to the case of maximized in-band and minimized out-of-band absorption values, which in the ideal case should converge to the step-function type absorption spectrum (Fig. 2b). Figure 3b shows the convergence plot of the TO FOM, illustrating the material evolution inside the unit cell. The spikes of the FOM convergence plot occur due to the applied filtering algorithm. Some of the generated 200 designs are shown in Fig. 3c, where the implemented filtering algorithm cuts off all sub-30-nm features. Here we only use the designs with: (i) more than 85% of average in-band and (ii) less than 40% of mean out-of-band absorption.

To be able to compare the performance of the generated designs, we have defined the efficiency of the thermal emitter as a product of in-band($eff^{in}$) and out-of-band($eff^{out}$) efficiencies. $eff^{in}$ is an in-band radiance of the emitter normalized to the in-band radiance of ideal emitter at 1800 C, while out-of-band efficiency $eff^{out}$ is defined as a ratio of the out-of-band radiance of back reflector and radiance of the TO design. The later reflects the fact that the response of the gap plasmon structures in the long-wavelength limit is fully determined by the material properties of the back reflector and limited by the optical losses of TiN. More details on the efficiency calculation can be found in Supplementary Materials, section S3. Figure 3d depicts the statistics of the design efficiencies. The normalized efficiency of the cylindrical emitter is 83%, while the best TO design has 92% efficiency. Figure 3e shows the absorption/emissivity spectra of the best TO (black) and cylindrical (red) designs. The obtained TO emitter designs with non-trivial shapes enable more uniform, higher in-band absorption while providing the same rapidly decaying tail in the out-of-band region. We note that the plasmonic back reflector entirely defines the behavior of the out-of-band tail of the absorption spectrum at longer wavelengths. Figure 3f shows the corresponding emission spectra of the best TO thermal emitter (black), cylindrical (red) vs. the black body emission spectrum at 1800 °C. A dense population of in-band modes in the TO design enables higher in-band emission that in the ideal case should match the black body in-band emission. The out-of-band emission is substantially suppressed in both design cases and is equal to 32% (TO design) and 29% (cylindrical emitter) of the out-of-band black-body radiation at 1800 C. The performance of the thermal emitter can be further optimized by adjusting optical properties of the back-reflector material, i.e. increasing the reflectivity and decreasing attenuation at long wavelength range.

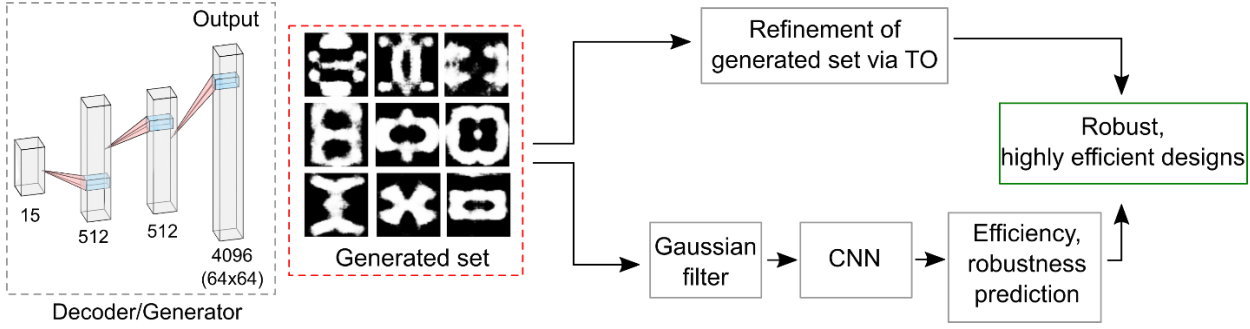

**Figure 4. Structure refinement and filtering schemes.** The generated design sets are refined via two different approaches: (i) by applying additional 15 iterations of TO and (ii) by applying the Gaussian filtering and passing through pre-trained convolutional neural network (CNN).

V. **AAE-OPTIMIZED THERMAL EMITTER AND LATENT SPACE ENGINEERING**

Once the AAE network is trained on the obtained TO designs, the encoder takes $64 \times 64$ binary, greyscale image/pattern of the TO design as input and compresses it into the 15-dimensional vector, representing a position in the 15-dimensional latent space. The decoder reconstructs the resonant layer design from the latent space by taking the latent coordinates as input. During the training process, the discriminator forces the encoder to form the latent space that would match the pre-defined distribution. In the initial case, we use Gaussian distribution as a pre-defined model.

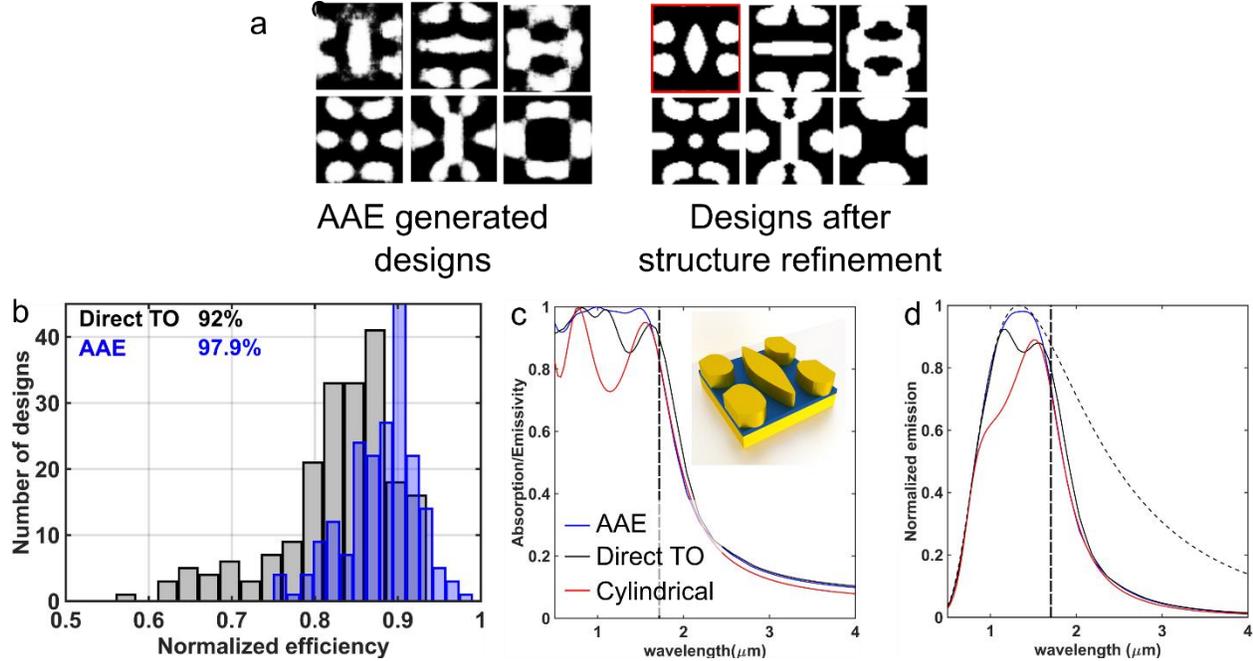

**Figure 5. AAE for design optimization of thermal emitters.** (a) Example of designs generated by the trained AAE (left panel); the same designs after the structure refinement process (right panel). (b) Statistics of the efficiency distribution for 200 designs obtained with AAE after the structure refinement process (blue bars). The same statistics of the 200-design set obtained with the direct TO (gray bars). (c) The absorption/emissivity spectra of the best AAE design in the set (blue curve), the best direct TO design (black), and the optimized cylindrical emitter (red). The inset shows the unit cell configuration of the best design in the set. (d) The corresponding emission spectra of the heaters at 1800 °C temperature. Black dashed curve corresponds to blackbody emission at 1800 °C.

Common to all deep neural networks, for efficient AAE training, a much larger designs dataset (~10k) is required compared with the available set of 200 designs. Since TO is time-consuming, the direct generation of thousands of designs is not practical. To overcome this issue, we used data augmentation, which takes into account the physical and symmetry properties of metasurfaces. Periodicity of the thermal emitter design allows for translational perturbation along a longitudinal direction without affecting the optical response of the structure. The freedom in cross-polarization selection allows for a 90-degree rotation. We use 20 random lateral translations of the original patterns and 90-degree rotated pattern per design that allowed us to enlarge the training set to 8400 samples. The AAE network is then trained on the expanded dataset. More details on the training process and specifics of the AAE structure are given in Supplementary Materials (section S3).

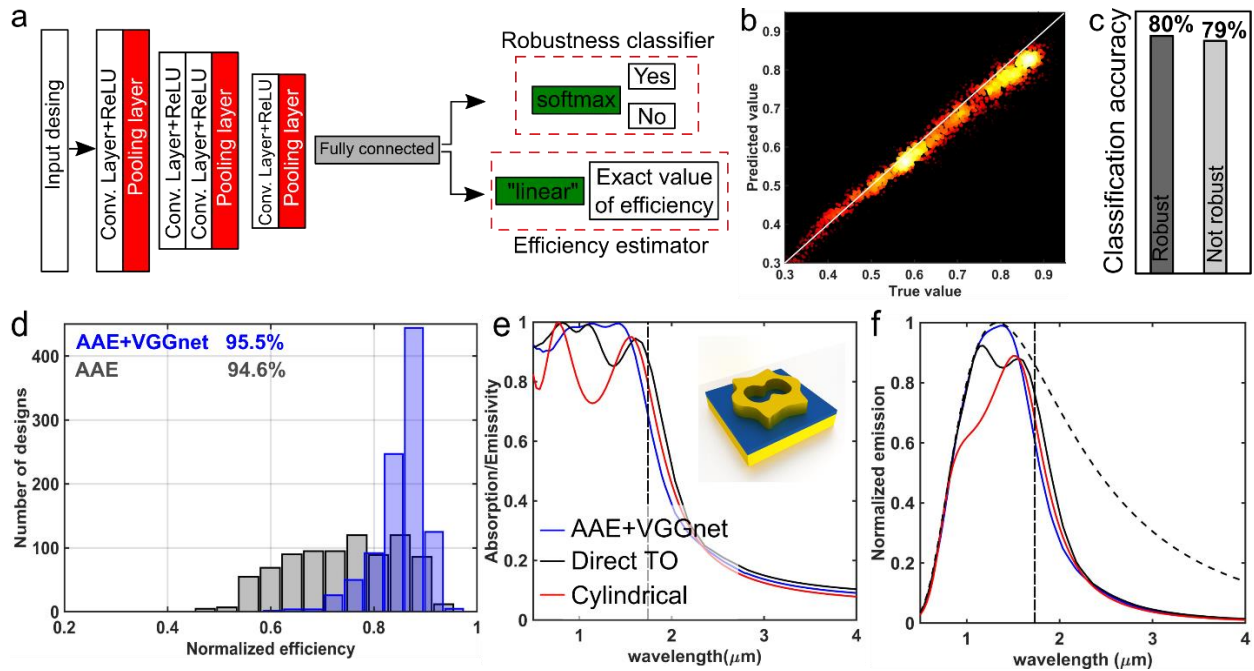

**Figure 6. AAE + VGGnet optimization.** (a) Schematics of the smaller VGGnet used for design efficiency prediction and robustness classification. (b) Regression map of the VGGnet performance on the training set. (c) Test results of the VGGnet based robustness classification scheme. (d) Statistics of the efficiency distribution for 1000 designs obtained directly from AAE network(gray bars) and AAE coupled with VGGnet filtering(blue bars). (e) The absorption/emissivity spectra of the best AAE+VGGnet design in the set (blue curve), the best direct TO design (black), and the optimized cylindrical emitter (red). The inset shows the unit cell configuration of the best design in the set. (f) The corresponding emission spectra of the heaters at 1800 °C temperature. Black dashed curve corresponds to blackbody emission at 1800 °C.

After the training phase, the decoder is used as a generator of new designs. The generated designs are refined/filtered with two different approaches (Fig.4): (i) using additional iterations of TO and

(ii) using a pre-trained convolutional neural network for predicting the efficiency and robustness of generated designs.

**Refinement within TO scheme.** The structure refinement with additional TO process ensures the stability of the final designs, as well as helps with eliminating sub-30 nm features. Using the decoder, we generate 1000 designs with a condition of having at least a 40% TiN filling factor within the domain. This requirement helps to prevent low-efficiency of the designs after the refinement due to a low plasmonic material fraction in the domain. The refinement uses the same constraints on the final design efficiency as the direct TO, i.e., (i) at least 85% of mean in-band absorption and (ii) less than 40% of the out-of-band (Fig. 5a.) Figure 4a indicates that the refinement removes the material "blurring", transforming the design into a binary structure. The AAE optimization provides a mean efficiency of 90% for the set of 200 designs vs. 82% for the set obtained with the direct TO (Fig. 5b). The best AAE designs (red box in Fig. 5a) provide the top efficiency of 98%, while the efficiency of the best pattern from the direct TO set is only 92%. The AAE design exhibits almost unit-level in-band absorption while having the same decaying tail of the out-of-band absorption spectrum as the direct TO and trivial emitters (Fig. 5c). The AAE designs enable almost all available in-band emission (98%) while significantly suppressing out-of-band radiation (30% of out-of-band black-body radiation at 1800 °C) (Fig. 5d).

**Pre-trained CNN based filtering.** Within the AEE+TO refinement optimization approach, most of the computational time is spent on structure refinement due to additional TO iterations. As an alternative, here we propose to use pre-trained CNN based filtering of highly-efficient and robust designs within AAE generated design set. CNN based structure filtering process consists of two main steps: (i) applying Gaussian filtering and binarization function to AAE generated design for elimination sub-30 nm feature (see Supplementary Materials, section S2) and (ii) robustness and efficiency estimation based on pre-trained CNN. Specifically, the CNN architecture utilized here is a smaller version of the VGGnet network, introduced in [60] (Fig. 6a). CNN takes 64 by 64 image of the design as an input and passes it through three hidden layers, which consist of convolutional layers with ReLU activation functions. Each hidden layer is followed by the max. pooling layer, which ensures the down-sampling of the feature maps. The stack of convolutional layers is followed by one fully-connected layer. The base VGGnet architecture is followed by two different activation functions of the final layer: (i) the "soft-max" with "cross-entropy" loss function for the robustness classification ("robust" or "not robust") and (ii) "linear" activation function with "mean squared error" loss function for efficiency prediction(regression). More details on the smaller VGGnet network can be found in the Supplementary Materials.

The VGGnet has been trained on 5000 designs generated by trained AAE on the original dataset. The efficiency and robustness ground truth labels have been assessed via FDTD simulation of each AAE generated designs. 70% and 30% of the dataset have been used for training and testing respectively. Fig. 6b shows the performance of the trained VGGnet regression model. The heatmap shows the statistics of the efficiency value prediction as a function of true labels. Here we can see that the VGGnet is able to predict efficiency with high accuracy. The coefficient of determination ($r^2$ coefficient) is equal to 0.93, which in the ideal case should be equal to 1. To be able to set up the robustness classification problem, we have labeled the design as a "robust" if $F = \max\left(\left|\text{eff}_{ideal} - \text{eff}_{eroded}\right|/\text{eff}_{ideal}, \left|\text{eff}_{ideal} - \text{eff}_{dilated}\right|/\text{eff}_{ideal}\right) \geq 0.95$, and as a "not robust" if $F < 0.95$, here $\text{eff}_j$ are efficiency values for ideal, eroded and dilated designs. Here we have applied $\pm 10\,nm$ perturbation of the ideal structure. Fig. 6c shows the test results for the robustness classification. Here we can see that VGGnet ensures 80% accuracy of "robust" devices classification and 79% for "not robust".

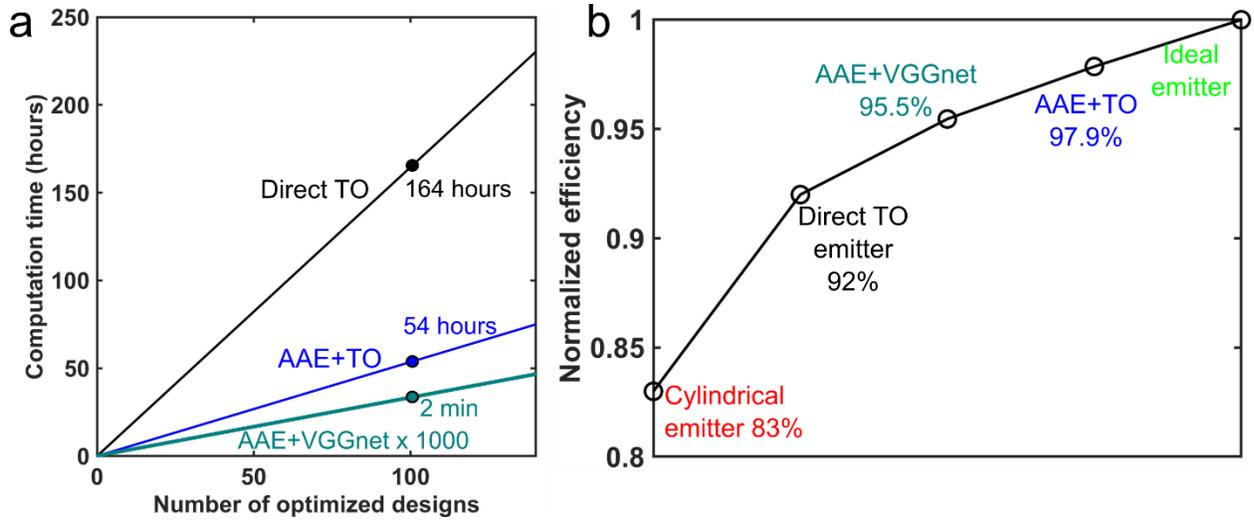

**Figure 7. Optimization search efficiencies.** (a) Dependence of the computational time of the direct TO (black), AAE+TO based optimization (blue) and AAE+VGGnet(cyan) on the number of the optimized high-efficiency resonant patterns. (b) Comparison of the efficiencies of the obtained best designs for all methods presented in this work.

Once both variations of VGGnet are trained, we have generated 1000 designs using the AAE+VGGnet approach with the two main constraints: high-efficiencies (>80%) and robustness of the design. Fig. 6d shows the statistics of the efficiency distribution of the generated AAE+VGGnet designs (blue bars) and 1000 designs generated directly from AAE (gray bars). For the elimination of the sub-30 nm features, AAE designs have been passed through the Gaussian filter. As we can see, the design set generated directly from AAE has almost the same distribution as a TO training set, while additional constrained filtering using VGGnet ensures high efficiency of the generated set. Almost 88% of the AAE+VGGnet generated designs have efficiency over 80%. The best AAE+VGGnet design has 95.5% efficiency, while the best design within the AAE set has 94.6%. The absorption spectrum, as well as emissivity spectrums of the best AAE+VGGnet generated design, are shown in Fig. 6d-e. For the sake of comparison spectrums of the best TO as well as cylindrical emitter, designs are shown as well.

Along with the higher efficiency, AAE assisted optimization ensures faster optimization search in comparison with conventional TO. Fig. 7a shows the comparison between computational costs of direct TO (black), AAE+TO refinement (blue) and AAE+VGGnet (red) approaches. Here we can see that for the generation of 100 designs direct TO requires 164 hours, while the AAE+TO refinement approach needs 54 hours. The direct TO needs 1.64 hours per design optimization on average, while the AAE based optimization needs only 31 min. The AAE optimization time consists of the decoder generation time (<1 s per design) and the refinement time (~31 min). In comparison, the AAE+VGGnet approach requires only 2 min for the generation of highly efficient (>80%) 100 designs, which is 1620 times faster than the AAE+TO approach and more than 4900 times faster than direct TO. All numerical simulations are done on a cluster node with two 12-core Intel Xeon Gold "Sky Lake" processors (24 cores per node) and 96 GB of memory. Direct full-wave simulation at each iteration is done in parallel, while the filtering, calculation of gradients and material distribution updates are performed in a sequential manner. Figure 7e depicts the results of all used optimization methods within this work. Here we would like to outline the main difference between two AAE assisted optimization approaches. The AAE optimization followed by additional TO refinement ensures the best solution and almost maximum possible result of the problem under consideration. However, TO based refinement procedure substantially limits the proposed approach in terms of computational time required. While the AAE+VGGnet approach ensures tremendous speed-up of the optimization search and relatively high overall device efficiency.

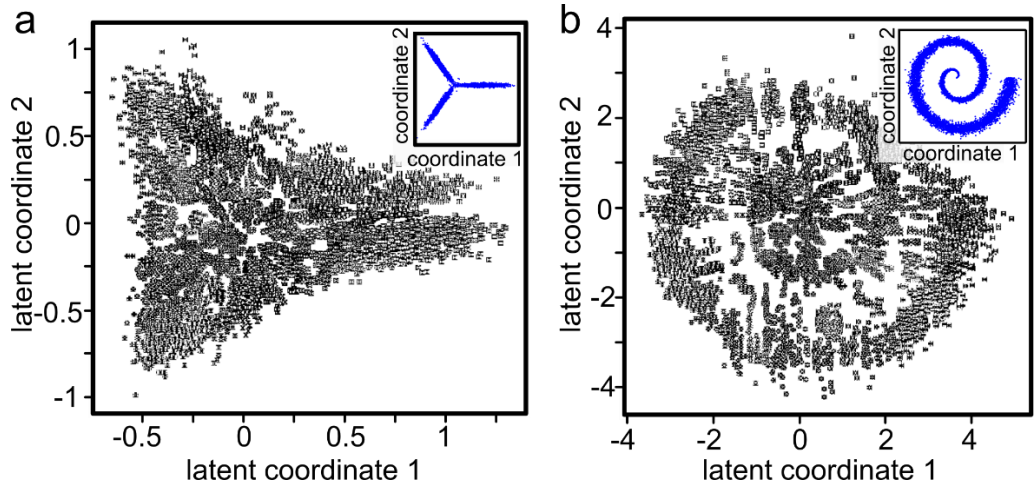

**Figure 8. Latent space engineering.** (a) 2D latent space distribution with "3 Gaussian mixture" sampling of the pre-defined model; (b) the same as in (a) but for "swiss roll" distribution. Insets show corresponding data sampling used as a pre-defined model.

The structure of the AAE network ensures unparalleled control over latent space distribution by adjusting the pre-defined model during the training phase. Such control can be adapted for the realization of global optimization techniques directly inside the compressed space. One possibility of realizing such global optimization is mapping the latent space distribution into the pre-defined surrogate model and use Bayesian optimization. Additionally, such control over the latent space could be used for the determination of sub-latent space with a lower dimension, corresponding to the best designs in the set by using principal component analysis. The analysis of such lower-dimensional spaces will allow determining the key requirement for achieving the best possible performance of the photonic/plasmonic device within a given design space. Within the aforementioned global optimization frameworks, the control over the configuration of the latent space is critical. Moreover, such latent space control is of significant importance for the multi-constrained problems, which require careful engineering of the compressed hyper-dimensional design spaces for the realization of efficient optimization searches.

To showcase such control, we train the AAE network on the TO design set along with probing it with the two types of pre-defined distributions: "3 Gaussian mixture"(Fig. 8a) and "Swiss roll" (Fig. 8b). "3 Gaussian mixture" is a mapping of normal random distribution into 3 2D Gaussian:

$$\tilde{x} = x\cos\left(\frac{2\pi}{3}\xi\right) - y\sin\left(\frac{2\pi}{3}\xi\right), \quad \tilde{y} = x\sin\left(\frac{2\pi}{3}\xi\right) + y\cos\left(\frac{2\pi}{3}\xi\right),$$

here $x, y$-a random number with normal distribution, $\xi$ random integer number with "discrete uniform" distribution between $[1,3]$. The "Swiss roll" distribution is a result of the mapping of random number with normal distribution into parametric 2D plane defined as:

$$\tilde{x} = 4\sqrt{\chi}\cos(4\pi\sqrt{\chi}), \quad \tilde{y} = 4\sqrt{\chi}\sin(4\pi\sqrt{\chi}), \quad \chi = 4(x+\xi),$$

here $x$-a random number with normal distribution, $\xi$ random integer number with "discrete uniform" distribution between $[1,3]$.

The AAE is trained to compress/reconstruct $64 \times 64$ input patterns into/from 2-D latent space. "3 Gaussian mixture" sampling indicates the formation of the three-lobe distribution of the design space, enforced by the pre-defined model distribution(Fig.8a). In a more sophisticated case, the pre-defined model is a complex spiral-shape distribution that is also reconstructed by the AAE accordingly (Fig. 8b).

## VI. CONCLUSION

The synergy between the inverse design methods and advanced machine learning techniques opens up a new paradigm to address highly complex, multi-constrained problems. Here, we merge the adjoint-based topology optimization with the AAE network and demonstrate faster optimization searches and unparalleled control over the latent space configuration. The latter is crucial for the realization of efficient optimization over high dimensional parametric landscapes, that is required for the design of multiconstrained, multifunctional photonic devices. Specifically, we optimize the design of a thermal emitter metasurface with high-efficiency thermal emission reshaping. We show that AAE+TO optimization-based emitter designs enable thermal reshaping with efficiencies up to 98%. Along with the better efficiency, the proposed AAE+VGGnet approach demonstrates

~4900 times faster optimization search in comparison with a conventional direct TO method. The proposed method can transform the area of optical design as well as data-driven materials synthesis for a plethora of applications in photonics, optoelectronics, MEMS and biomedical synthetics.

## SUPPLEMENTARY MATERIAL

See the supplementary material for details on dielectric permittivity functions of TiN and $Si_3N_4$, additional information on topology optimization framework, adversarial autoencoder structure, training process, data augmentation procedure; additional data on AAE based optimized designs; structure of the VGGnet.

## ACKNOWLEDGMENTS

This work was supported in part by the AFOSR (FA9550-17-1-0243), NSF (0939370-CCF), DARPA/DSO Extreme Optics and Imaging (EXTREME) Program (HR00111720032).

# Supplementary Materials

# for

# Machine-Learning-Assisted Metasurface Design for High-Efficiency Thermal Emitter Optimization

Zhaxylyk A. Kudyshev[1,2], Alexander V. Kildishev[1],
Vladimir M. Shalaev[1], and Alexandra Boltasseva[1]

[1] School of Electrical and Computer Engineering and Birck Nanotechnology Center, Purdue University, West Lafayette, IN, 47906, USA
[2] Center for Science of Information, Purdue University, West Lafayette, IN, 47906, USA

**S1. DIELECTRIC PERMITTIVITY FUNCTIONS OF TiN AND $Si_3N_4$**

Dielectric permittivity functions are obtained by using a custom-built platform that comprises a heating stage integrated onto a spectroscopic ellipsometer setup. More information regarding the experimental setup and permittivity retrieval process can be found in[1]. Particularly, dielectric functions of TiN and Si3N4 have been used in the optimization process, correspond to 1000 C temperature response (Fig.S1).

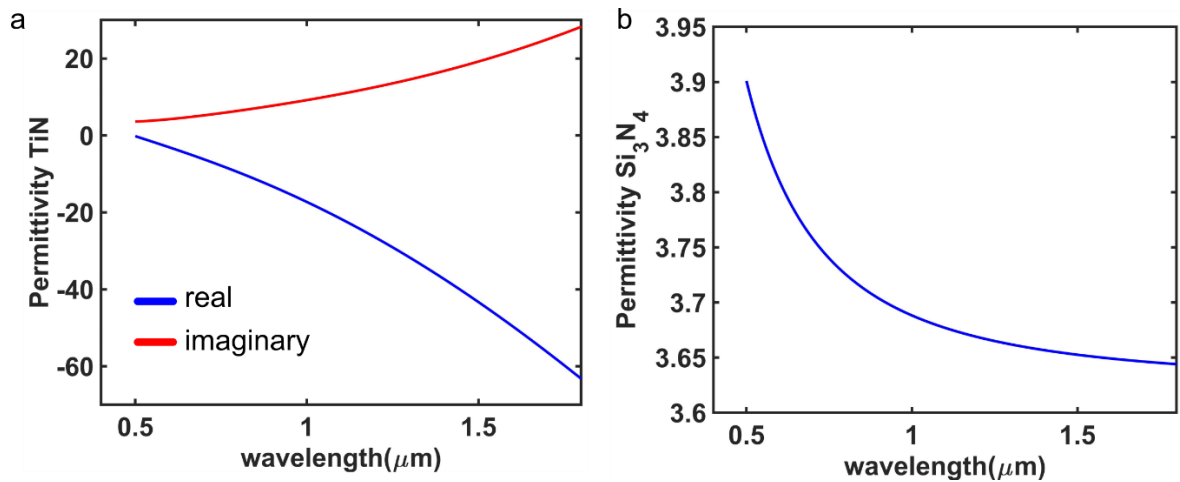

**Figure S1. Dielectric permittivity of TiN (a) and $Si_3N_4$ (b)**

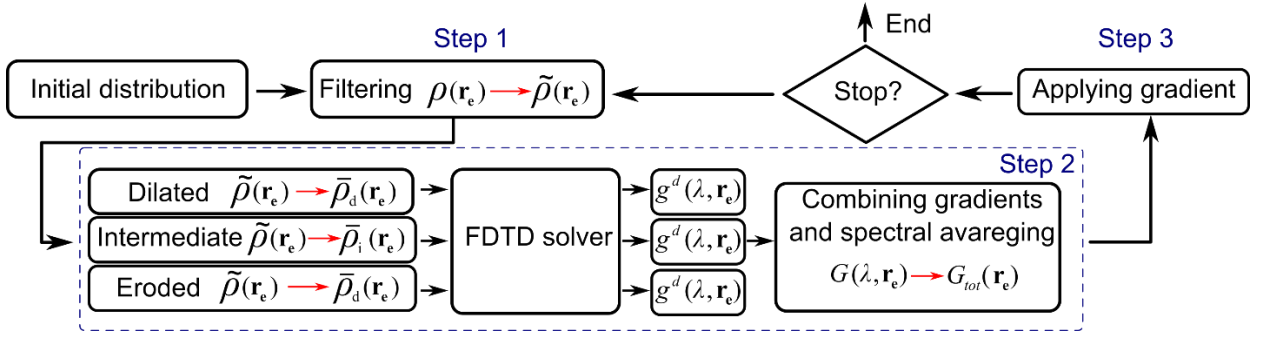

**Figure S2. The general flow of the direct TO**

## S2. TOPOLOGY OPTIMIZATION FRAMEWORK

For this study, we have adapted adjoint-based topology optimization (TO) techniques previously applied to photonic crystal structures [2], waveguide structure [3], and dielectric metagratings [4]. The TO is done using the Matlab scripting language as an application programming interface coupled with a commercial finite-difference time-domain (FDTD) direct solver (Lumerical FDTD). TO consists of three main steps: (i) *filtering*; (ii) *FOM gradient calculation*; (iii) *updating of the material distribution*. The general flow of direct topology optimization is shown in Fig. S2.

**Material interpolation.** The material in the optimization region is set to a smooth profile of a dielectric function, changing continuously from the dielectric permittivity of air to the dielectric permittivity of TiN. The material distribution at the $n^{th}$ iteration and at $\mathbf{r}_e$ location is defined by a continuous function $\rho^{(n)}(\mathbf{r}_e) \in [0,1]$. The wavelength-dependent permittivity distribution $\varepsilon_r$ of the TiN/air mixer is defined through a non-linear interpolation scheme, which is shown to be more applicable for TO of plasmonic structures [5]:

$$\varepsilon_r\left(\eta(\lambda,\rho^{(n)}),\kappa(\lambda,\rho^{(n)})\right) = \left(\eta(\lambda,\rho^{(n)})^2 - \kappa(\lambda,\rho^{(n)})^2\right) - i\left(2\eta(\lambda,\rho^{(n)})\kappa(\lambda,\rho^{(n)})\right), \quad \text{(S1)}$$

$$\begin{aligned}\eta(\lambda,\rho^{(n)}) &= \eta_{air} + \rho^{(n)}\left(\eta_{TiN}(\lambda) - \eta_{air}\right),\\ \kappa(\lambda,\rho^{(n)}) &= \kappa_{air} + \rho^{(n)}\left(\kappa_{TiN}(\lambda) - \kappa_{air}\right)\end{aligned} \quad \text{(S2)}$$

with $\eta_j(\lambda) = \sqrt{(|\tilde{\varepsilon}_j(\lambda)| + \Re(\tilde{\varepsilon}_j(\lambda)))/2}$, $\kappa_j(\lambda) = \sqrt{(|\tilde{\varepsilon}_j(\lambda)| - \Re(\tilde{\varepsilon}_j(\lambda)))/2}$, $\tilde{\varepsilon}_j(\lambda)$ - permittivity function of TiN ($\tilde{\varepsilon}_{TiN}$) or air ($\tilde{\varepsilon}_{air}$), index j corresponds to the material component.

**Step 1: Filtering.** The elimination of the sub precision features during TO is done using a filtering technique [6,7]:

$$\rho^{(n)}(\mathbf{r_e}) = \frac{\sum_{j \in N_e} \left(R - \|\mathbf{r_j} - \mathbf{r_e}\|\right) v_j \rho^{(n)}(\mathbf{r_j})}{\sum_{j \in N_e} \left(R - \|\mathbf{r_j} - \mathbf{r_e}\|\right) v_j} \qquad (S3)$$

here $\mathbf{r_e}$ - location of the filtered element $e$, $N_e$ - the total number of elements around $e$ within the filtering area of radius $R$, $v_j$ - a volume of the element $j$.

**Step 2: FOM calculation.** The figure of merit (FOM) of TO is defined as a weighted average of the in-band absorption and out-of-band reflectivity. The spectral averaging is done with respect to the step-function type emissivity spectrum of the ideal emitter. The adjoint formalism allows to define the gradient of FOM at a given wavelength, $g(\lambda, \mathbf{r_e}) = \frac{\partial \text{FOM}(\lambda, \mathbf{r_e})}{\partial \varepsilon_r}$, as a product of field distributions inside the optimization domain of forward ($\mathbf{E}(\lambda, \mathbf{r_e})$) and adjoint ($\mathbf{E}^{adj}(\lambda, \mathbf{r_e})$) simulations [3,4]:

$$g(\lambda, \mathbf{r_e}) = \pm 2\omega^2 v_e \, \text{Re}\left(\bar{r}(\lambda) \mathbf{E}(\lambda, \mathbf{r_e}) \cdot \mathbf{E}^{adj}(\lambda, \mathbf{r_e})\right) \qquad (S4)$$

here $\bar{r}(\lambda)$ is the conjugate of the complex reflection coefficient. Minus sign is used for the in-band part ($\lambda_{min} \leq \lambda \leq \lambda_{max}$), while the plus is used for the remaining out-of-band region ($\lambda \geq \lambda_{max}$).

**Robustness control.** The robustness control is done by applying the threshold projection to $\tilde{\rho}^{(n)}(\mathbf{r_e})$ as:

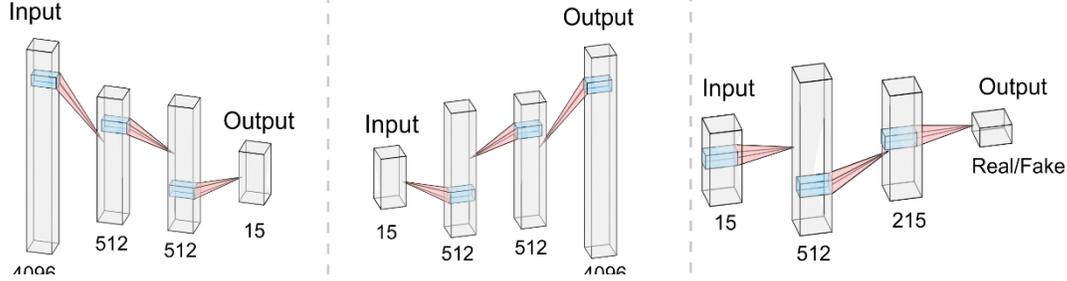

**Figure S3. Structure of the AAE components**

$$\bar{\rho}^{(n)}(\mathbf{r_e}) = \begin{cases} \eta\left(\exp\left[-\beta(1-\tilde{\rho}^{(n)}/\eta)\right] - \exp[-\beta](1-\tilde{\rho}^{(n)}/\eta)\right), & 0 \leq \tilde{\rho}^{(n)} \leq \eta \\ (1-\eta)\left(1-\exp\left[-\beta(\tilde{\rho}^{(n)}-\eta)/(1-\eta)\right]\right) + \exp[-\beta](\tilde{\rho}^{(n)}-\eta)/(1-\eta), & \eta < \tilde{\rho}^{(n)} \leq 1 \end{cases} \quad (S5)$$

Three threshold values are defined such as $0 < \eta_d < \eta_i < \eta_e \leq 1$, giving three cases for the design: (i) dilated ($\eta_d < 0.5$), (ii) intermediate ($\eta_i = 0.5$), (iii) eroded ($\eta_e > 0.5$). $\beta$ is a parameter of the threshold function sharpness. This parameter is increased during the TO process to gradually push the material distribution inside the optimization domain to become a binary composition.

All three designs are passed to the direct solver, which returns corresponding field distributions of the forward ($\mathbf{E}(\lambda, \mathbf{r_e})$) and adjoint ($\mathbf{E}^{adj}(\lambda, \mathbf{r_e})$) simulations. Using Eq. S4, three FOM gradient profiles $g^d(\lambda, \mathbf{r_e})$, $g^i(\lambda, \mathbf{r_e})$, $g^e(\lambda, \mathbf{r_e})$ are calculated.

The resulting FOM profile at a given wavelength is defined as:

$$G^{(n)}(\lambda, \mathbf{r_e}) = \sum_q \sum_{j \in N_e} g^q(\lambda, \mathbf{r}_j) \frac{\partial \bar{\rho}_q^{(n)}(\mathbf{r}_j)}{\partial \tilde{\rho}^{(n)}(\mathbf{r}_j)} \frac{\partial \tilde{\rho}^{(n)}(\mathbf{r}_j)}{\partial \rho^{(n)}(\mathbf{r_e})} \quad (S6)$$

where index $q$ corresponds to the perturbed design type: dilated ($q = d$), intermediate ($q = i$), or eroded ($q = e$). The first derivative in (S6) is determined by (S5), while the second one is determined through (S3).

The derivatives in (S6) are calculated with the help of (S3) and (S5). The spectral averaging is used to determine the final gradient that is applied to the base design pattern:

$$G_{tot}^{(n)}(\mathbf{r_e}) = \sum_{\lambda} \begin{cases} |r(\lambda)|^2 G^{(n)}(\lambda, \mathbf{r_e}), & \lambda_{min} \leq \lambda \leq \lambda_{max} \\ \left(1 - |r(\lambda)|^2\right) G^{(n)}(\lambda, \mathbf{r_e}), & \lambda \geq \lambda_{max} \end{cases} \quad (S7)$$

**Step 3: Material distribution update:** Once the total FOM gradient is determined, the material distribution function is updated as:

$$\rho^{(n+1)}(\mathbf{r_e}) = \rho^{(n)}(\mathbf{r_e}) + c G_{tot}^{(n)}(\mathbf{r_e}) \quad (S8)$$

Steps (1)-(3) are repeated for 50 iterations. The filtering is applied at each 10$^{th}$ iteration. The final design is saved once it meets the main optimization constraints.

## S3. EFFICIENCY CALCULATION

The emission spectrum is determined by its spectral emissivity $\varepsilon(\lambda)$ of the emitter as:

$$I(\lambda, T) = \varepsilon(\lambda) B_\lambda(\lambda, T),$$

here $B_\lambda(\lambda, T) = 2h v^3 / \left(c^2 \left(\exp(h v / k_B T) - 1\right)\right)$ - the spectral radiance of the black body at a given temperature $T$, $v$ is a frequency, $h$ - Planck constant, $k_B$ - Boltzmann constant, $c$ - the speed of light in free space.

The radiance of the thermal emitter within the spectral range of interest ($\lambda \in [\lambda_{min}, \lambda_{max}]$) can be by integrating the emission spectrum over the wavelength:

$$Q(T) = \int_{\lambda_{min}}^{\lambda_{max}} I(\lambda, T) d\lambda.$$

The ideal thermal emitter has a step function type emissivity function shown on (Fig. 2b, main text). The requirement on the in-band emissivity spectrum can be addressed by the careful engineering corresponding resonant response of the top layer antenna. However, the response of the gap plasmon structures in the long-wavelength limit is fully determined by the material properties of the back reflector. To take into account this fact into the emitter specs, the efficiency of the thermal emitter is set to be the product of the in-band and corrected out-of-band efficiencies as:

$$eff = eff^{in} \cdot eff^{out},$$

here

$$eff^{in} = \frac{\int_{\lambda_{min}}^{\lambda_{max}} \varepsilon(\lambda) B_\lambda(\lambda,T) d\lambda}{\int_{\lambda_{min}}^{\lambda_{max}} B_\lambda(\lambda,T) d\lambda}, \qquad eff^{out} = \frac{\int_{\lambda_{max}}^{\infty} \varepsilon_{TiN}(\lambda) B_\lambda(\lambda,T) d\lambda}{\int_{\lambda_{max}}^{\infty} \varepsilon(\lambda) B_\lambda(\lambda,T) d\lambda}.$$

$\varepsilon(\lambda)$, $\varepsilon_{TiN}(\lambda)$ - spectral emissivity of the optimized emitter and bare TiN back reflector, $T$ - the working temperature of the emitter, $\lambda_{min}, \lambda_{max}$ - lower and upper bounds of the PV cell's spectral working band.

## S4. ADVERSARIAL AUTOENCODER FOR DESIGN PRODUCTION

**Structure of AAE.** AAE consists of three coupled neural networks: the encoder, decoder/generator, and discriminator [8]. Figure S3 shows a detailed description of the neural networks.

*Encoder:* The encoder takes a 4096-dimensional vector (that corresponds to a 64 × 64 binary design pattern) as an input. We use two fully-connected layers as the hidden layers of the encoder and a 15 neuron as an output layer of the encoder so that each of the hidden layers has 512 neurons.

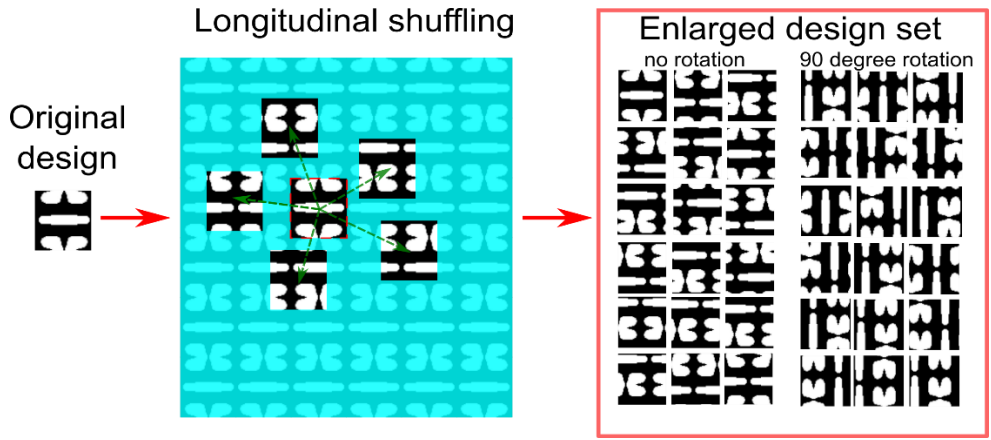

**Figure S4. Data augmentation of the TO design set**

For hidden layers, the rectified linear unit (ReLU) activation function is used, and one batch normalization layer is coupled to the second linear layer.

*Decoder:* The decoder has the same architecture as the encoder but with the reversed sequence. The decoder generates a 4096-element output vector based on 15-dimensional input. For the output layer, we use tanh activation function.

*Discriminator*: The discriminator takes a 15-dimensional latent vector as an input and binary perform classification (fake/real), so the output is one neuron. Here we have used 2 hidden liner layers with 512 and 256 neurons. The activation function for two hidden layers is the ReLU and for the output layer is the sigmoid function.

**Data augmentation.** Data augmentation, which takes into account the physical and symmetry properties of meta-device, is used to expand the training set of the problem. Figure S4 shows the schematics of the data augmentation process. Due to the translational symmetry of the thermal emitter design, it is possible to stack a single TO design into a continuous pattern. Gradually scanning this periodic super-pattern with a $280 \times 280$ nm$^2$ window in both lateral directions makes it possible to generate different "versions" of the same design.

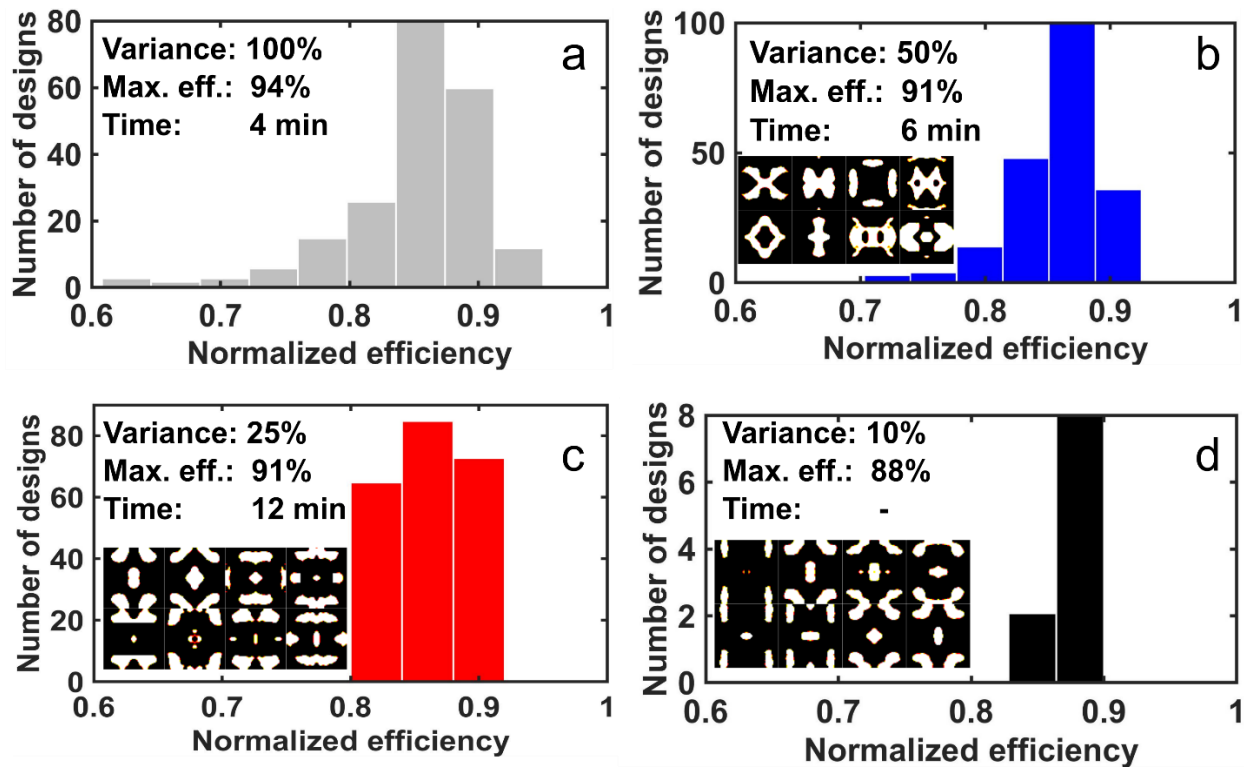

**Figure S5. Data variance.** Statistics of generated dataset on variance level: (a) 100%, (b) 50%, (c) 25% and (d) 10%. Inset shows the efficiency of best design in the set, time required for generating 200 designs. In the case of 10% data variance, AAE failed to generate >10 designs with pre-defined efficiency and robustness within 1 hour. Samples of the generated designs are shown in the inset pictograms.

Along with this, we use a 90-degree rotation to double the set of available resonant patterns. The latter is possible due to the freedom in choosing the direction of the primary polarization axis. We use 20 random lateral translations of the original pattern doubled with a 90-degree rotated pattern per design, significantly expanding the training set (up to 8400 resonant patterns).

**Variance of the training data.** With the data augmentation technique, one of the important questions that should be highlighted is an influence of the variance of the augmented training set to the AAE performance. To be able to study this question in more detail we have trained the AAE network on the datasets with the same size but with different variance levels of the designs. For this purpose four datasets with the same total size (8400 designs) have been formed:

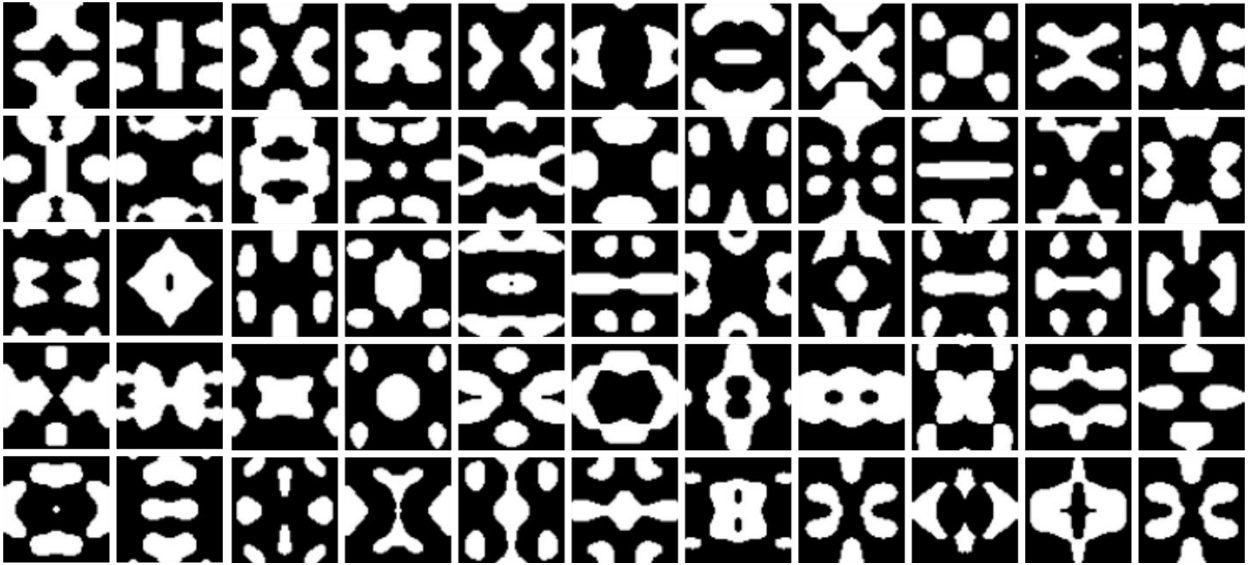

**Figure S6. The top 55 AAE-optimized resonant patterns.** White color corresponds to TiN, black color corresponds to air.

- 100% variance: 200 TO designs with two rotation (0, 90 degrees) and 20 random lateral translations of the original pattern (original dataset);
- 50% variance: best 100 TO designs with two rotation (0, 90 degrees) and 40 random lateral translations of the original pattern;
- 25% variance: best 50 TO designs with two rotation (0, 90 degrees) and 80 random lateral translations of the original pattern;
- 10% variance: best 20 TO designs with two rotation (0, 90 degrees) and 200 random lateral translations of the original pattern.

Figure S5 shows the main result of the analysis for all four cases. In all cases we have used VGGnet to filter out robust and high efficient (>80% efficiency) designs and targeted to generate 200 designs in total. The corresponding time required to generated constrained 200 designs are indicated on corresponding figures. Noting that in the case of 10% data variance AAE has generated only 10 robust, high efficient designs in one hour. The statistics of the efficiency distribution in the generated sets show narrower distribution with decreasing the variance in the training set. With decreasing variance of the training set, AAE becomes biased toward designs with curtain shaped and efficiencies, and since we have used only best TO designs (efficiencies in between 80% and 90%), resulted efficiency distributions are localized around this region. Such shape biasing is clearly seen from the design samples for 10% and 25% variance cases. The 10% case shows the bias mainly to two different shapes.

**Training phase.** Training of the AAE consists of two phases. The first phase aims to minimize the reconstruction error of the encoder-decoder system. In this step, the input pattern is passed through the encoder to get the latent vector, which is then used as the decoder input. The decoder produces the output pattern. So the backpropagation method is used to update the weights of the encoder/decoder nets to minimize the difference between the input and output patterns. The second phase aims to set the adversarial feedback in the network by coupling the discriminator with the encoder. The discriminator is trained to classify between output the encoder and the random input with a predefined distribution. Once the discriminator is trained, it is then used to update the encoder to minimize adversarial loss applied to the latent vector distribution of the encoder output.

## S5. AAE-OPTIMIZED DESIGNS OF HIGH-EFFICIENCY THERMAL EMITTERS

Figure S6 shows design profiles of top 55 AAE-optimized resonant patterns, with the unit cell area of $280 \times 280$ nm$^2$.

## S6. COMPARISON OF AAE AND GAN NETWORKS

The choice of the AAE network over other deep generative network architectures, such as VAE and GAN, is motivated by several, key advantages, such as (i) dense latent space distribution and (ii) un-parallel control over the latent space configuration.

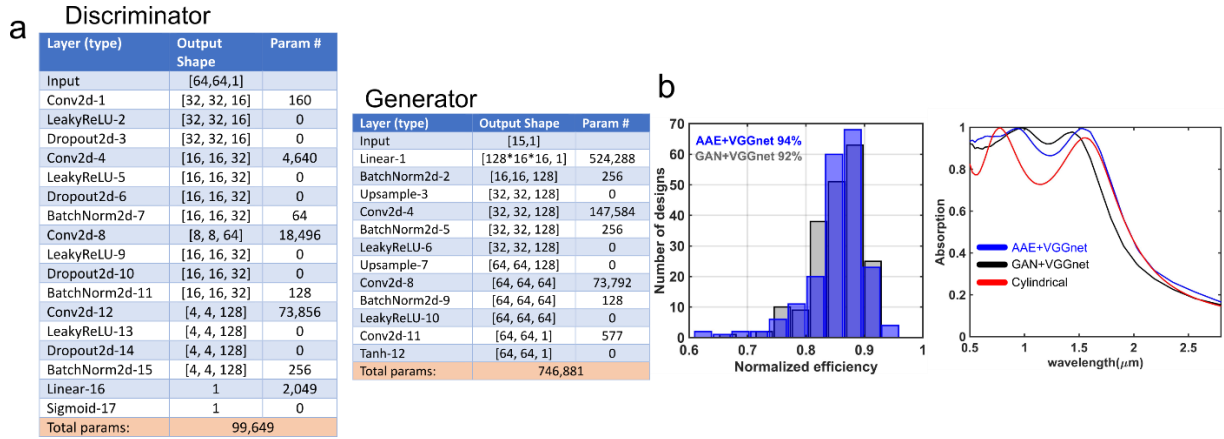

**Figure S7. GAN vs AAE comparison.** (a) Structure of the GAN network. (b, left) Statistics of generated dataset byAAE+VGGnet (blue) and GAN+VVGnet(gray); (b, right) emissivity spectrums for the best GAN(black) and AAE(blue) designs.

Here we have performed a quantitative comparison of the performance of GAN and AAE networks trained on topologically optimized design set used in the main text. The structure of the discriminator and generator of the GAN is shown in Fig. S7a. Once the generator of the GAN network was trained, it has been used to generate 200 design. The GAN network has been coupled with VGGnet for filtering robust and efficient(>80%) designs. The comparison between statistics of the efficiency distributions of GAN+VGGnet and AAE+VGGnet networks is shown in Fig. S7b(left). The best AAE+VGGnet design in the generated set shows 94% efficiency, while the best design generated by GAN has 92% efficiency. Corresponding emissivity spectrums are shown in Fig. S7b(right). Noting that the best design in the training has 92% efficiency as well. In the main text, we have shown that AAE+VGGnet (1000 design set) shows 95.5%, while AAE+TO shows 98%. The GAN network required 29 min for generated 200 designs with predefined robustness and efficiency (>80%), while the AAE network requires only 4 min. All this shows that the AAE based approach shows better performance in terms of the best design efficiency and time requirements.

| Layer (type) | Output Shape | Param # |
|---|---|---|
| Input | [64,64,3] | |
| conv2d_1 | [64, 64, 32] | 896 |
| activation_1 | [64, 64, 32] | 0 |
| batch_normalization_1 | [64, 64, 32] | 128 |
| max_pooling2d_1 | [21, 21, 32] | 0 |
| dropout_1 | [21, 21, 32] | 0 |
| conv2d_2 | [21, 21, 64] | 18496 |
| activation_2 | [21, 21, 64] | 0 |
| batch_normalization_2 | [21, 21, 64] | 256 |
| conv2d_3 | [21, 21, 64] | 36928 |
| activation_3 | [21, 21, 64] | 0 |
| batch_normalization_3 | [21, 21, 64] | 256 |
| max_pooling2d_2 | [10, 10, 64] | 0 |
| dropout_2 | [10, 10, 64] | 0 |
| conv2d_4 | [10, 10, 128] | 73856 |
| activation_4 | [10, 10, 128] | 0 |
| batch_normalization_4 | [10, 10, 128] | 512 |
| conv2d_5 | [10, 10, 128] | 147584 |
| activation_5 | [10, 10, 128] | 0 |
| batch_normalization_5 | [10, 10, 128] | 512 |
| max_pooling2d_3 | [5, 5, 128] | 0 |
| dropout_3 | [5, 5, 128] | 0 |
| flatten_1 | [3200] | 0 |
| dense_1 | [1024] | 3277824 |
| activation_6 | [1024] | 0 |
| batch_normalization_6 | [1024] | 4096 |
| dropout_4 | [1024] | 0 |
| dense_2 (Regression model) | [4] | 4100 |
| dense_3 (Regression model) | [1] | 5 |
| Total params: | | 3,565,449 |
| dense_2 (Classification model) | [2] | 2050 |
| activation_7 (Classification model) | [2] | 0 |
| Total params: | | 3,563,394 |

**Figure S8. VGGnet structure.**

**S8. VGGNET STRUCTURE** CNN takes 64 by 64 image of the design as an input and passes it through three hidden layers, which consist of convolutional layers with ReLU activation functions. Each hidden layer is followed by the max. pooling layer, which ensures the down-sampling of the feature maps. The stack of convolutional layers is followed by one fully-connected layer. The base VGGnet architecture is followed by two different activation functions of the final layer: (i) the "soft-max" with "cross-entropy" loss function for the robustness classification ("robust" or "not robust") and (ii) "linear" activation function with "mean squared error" loss function for efficiency prediction(regression). A detailed description of the VGGnet is shown in Fig. S8. Both classification and regression models use the same base architecture of the VGGnet type network, The main difference between two networks is in the final layers (outlined by red boxes).

**S9. AAE OPTIMIZATION FOR DIFFERENT UNIT CELL DIMENSIONS**

To be able to test the performance of the proposed AAE based optimization approach we have trained the AAE network on TO designs sets corresponding to two different unit cell sizes: 250 nm and 300 nm. For each of the unit cell sizes, we have optimized 150 designs with the same optimization scheme and parameters used for 280 unit cell case. The TO datasets have been enlarged using the same augmentation scheme and resulted in 8400 designs per unit cell configuration. The AAE network has been coupled with pre-trained VGGnet for filtering robust and highly efficient designs(>80%).

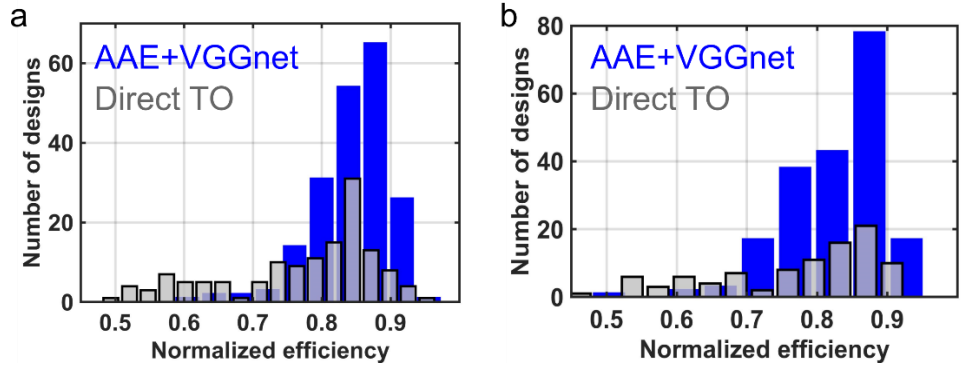

**Figure S9. AAE for different unit cell size.** Statistics of efficiency distribution of generated designs using direct TO (grey) and AAE+VGGnet approach(blue) for (a) 250 nm unit cell and (300 nm unit cell sizes.)

Fig.S9 shows the statistics of the generated 200 designs for 250 nm unit cell size (a) and 300 nm (b). From this comparison, it can be seen that the AAE+VGGnet approach in both cases generates the design set with higher mean efficiency in comparison with direct TO and better performance of the best designs in the set. Specifically, the best designs for the 250 nm unit cell case ensure 95.75% (AAE+VGGnet) and 94.5% (direct TO). For the case of 300 nm unit cell, the best design generated by AAE+VGGnet ensures 95% while direct TO ensures only 92%. Moreover, the AAE+VGGnet approach ensures the generation of 200 highly efficient designs within 4 min, while direct TO requires 328 for the generation of 200 TO designs.